\documentclass[showpacs,aps,pra,floatfix,reprint,superscriptaddress,footinbib,citeautoscript]{revtex4-2} 

\usepackage{amssymb,amsfonts,amsmath}
\usepackage{graphicx}
\usepackage{verbatim}
\usepackage[utf8]{inputenc} 
\usepackage[T1]{fontenc} 
\usepackage{soul,xcolor,colortbl}
\usepackage{multirow}
\usepackage{bm} 
\usepackage{array}
\usepackage{upgreek}
\usepackage{chemmacros}
\usepackage{color} 
\usepackage{hyperref} 
\hypersetup{colorlinks=true,citecolor=blue,linkcolor=blue,urlcolor=blue} 
\usepackage{mdframed} 
\usepackage{longtable} 
\usepackage{enumitem}
\usepackage{booktabs} 
\usepackage{lipsum} 
\usepackage{orcidlink} 
\usepackage{makecell}

\newcolumntype{L}[1]{>{\raggedright\arraybackslash}p{#1} }
\newcolumntype{C}[1]{>{\centering \arraybackslash}p{#1} }
\newcolumntype{R}[1]{>{\raggedleft \arraybackslash}p{#1} }

\DeclareSIUnit\inch{inches}

\setcitestyle{square}

\makeatletter \renewcommand\frontmatter@abstractwidth{\dimexpr\textwidth\relax} \makeatother 

\def\MEMS{Department of Mechanical Engineering and Materials Science, Duke University, Durham, NC 27708, USA}
\def\CXM{Center for Extreme Materials, Duke University, Durham, NC 27708, USA}
\def\PHTAU{Department of Physical Electronics, Tel Aviv University, Israel}
\def\NRCN{Department of Physics, NRCN, Israel}

\begin{document}

  \title{Composition/structure directed search for new chalcogenide compounds}
\author{Alon~Hever}\affiliation{\PHTAU}%
\author{Ohad~Levy}\affiliation{\NRCN}\affiliation{\CXM}%
\author{Stefano~Curtarolo\,\orcidlink{0000-0003-0570-8238}}\affiliation{\CXM}\affiliation{\MEMS}
\author{Amir~Natan\,\orcidlink{0000-0003-4517-5667}}\affiliation{\PHTAU}%

\begin{abstract}
\noindent This work presents a simple scheme for finding new crystalline compounds by adapting structure types from neighbor atoms compounds. The approach is demonstrated for the selenide and sulfide families of binary compounds. It predicts ten new compounds that are not currently included in the inorganic crystal structure database (ICSD). The compounds primarily originated from a small search domain that includes near neighbors. Comparison with extended searches that include structures from binary systems of more remote atoms in the periodic table demonstrate the relative efficiency of near neighbor screening. This points at the possibility of using similar directed searches as a heuristic rule for efficiently finding new stable compounds in additional compound families.
\end{abstract}

\maketitle

\noindent

\section*{Introduction}
\label{sec:Introduction}

Predicting the existence of ordered structures in alloy systems from their components is a significant
challenge of current materials research~\cite{Maddox1988, Oganov2018}. It requires reliable evaluation and comparison of the
energies of competing spatial arrangements of elements in the same element set to identify those
with the lowest energies as stable or metastable compounds~\cite{Oganov2010}. First-principles calculations can address
the reliable evaluation of individual structures' energy within the framework of Density Functional
Theory (DFT)~\cite{Hohenberg1964, Kohn1965, Stanton2001}, which is currently the workhorse of computational 

materials science, and offers
the best balance of accuracy and computational cost. However, identifying stable compounds is more
complicated since it requires considering a large number of structures. As more degrees of freedom are added;
e.g., the number of elements and different stoichiometries; the number of potential phases
becomes too large for practical examination~\cite{Oganov2018, Oganov2010, Morgan2005, Curtarolo2013}. Efficient computational routes for discovering new
materials are essential to narrow the exploration of atom arrangements in 3D space, which severely
limits brute force trial and error approaches. To overcome this limitation, stability determination
procedures employ a medley of selection methods that reduce the number of DFT calculations.

In searching for new stable compounds, one can either rely on naturally occurring structures as an
initial guess or use various algorithms that do not need prior structure data to generate new
structures. Some of the large high throughput database projects, such as AFLOW~\cite{Curtarolo2012}, the Materials
Project (MP)~\cite{Jain2013}, and the Open Quantum Materials Database (OQMD)~\cite{Saal2013}, use naturally occurring
structures as the starting point for new compounds search. This is usually based on the initial
consideration of input structures included in the extensive empirical inorganic crystalline structure
database (ICSD)\cite{Hellenbrandt2004}. It has been demonstrated that these methods are capable of identifying
previously unsuspected stable compounds~\cite{Curtarolo2013, Hautier2010, Hautier2012, Montoya2020, Zhang2019, Huang2018}, and lead to the experimental discovery
of new compounds~\cite{Oganov2018, Jain2016}. Examples of methods that do not require prior structure data are evolutionary
algorithms~\cite{Lyakhov2013, Oganov2006} and random structure search~\cite{pickard2011ab}. However, even these algorithms usually employ
naturally occurring structures as initial seed trials and may be significantly
accelerated by adding them to the search.

The efficiency of the process depends on the input structures selected in the first step. The ICSD
database contains about 318,901 entries (as of March 2025) of crystalline materials entries. The
entries in the ICSD are not necessarily unique since several entries can represent the same compound
from different experiments. Compounds of different elements arranged in the same crystallographic
positions are said to have the same structure type---e.g., both NaCl and NiO can adopt the rock salt
structure type. Usually, those structure types are characterized by their stoichiometry, Pearson
symbol, and space group. There are currently more than 13,000 such prototypes in the ICSD.
Obviously, the initial structure selection can not possibly carry out a comprehensive survey of all
structure types, even as seeds for advanced machine learning algorithms. Since the search for new potential
compounds in any specific alloy system may miss relevant structure types, narrowing the
space of these relevant structure types becomes crucial to streamline the search and provide a
manageable basis for a more efficient survey of potential new compounds.

\begin{figure*}[htb!]
    \centering\includegraphics[width=\textwidth]{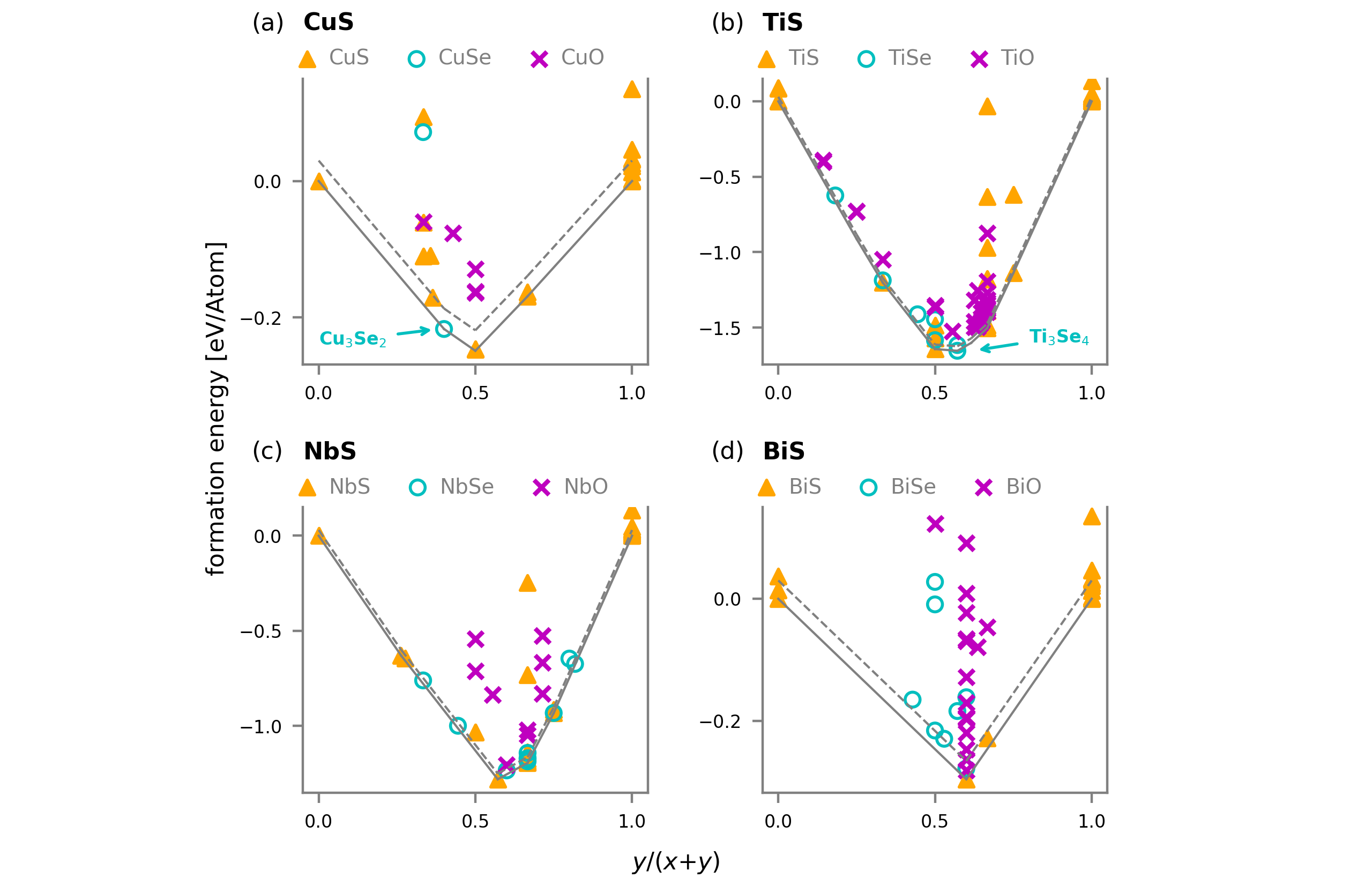}
    \caption{\small Convex hulls of the binary sulfide ($A$S) systems
      enriched with the structure types of the corresponding selenide
      ($A$Se) and oxide ($A$O) compounds. The orange triangle markers
      represent the known compounds listed in ICSD for each system
      (including the pure phases of the elements). The enriching
      structure types from the corresponding Se and O systems are
      shown in cyan circles and magenta $x$-symbols, respectively. Bold
      lines indicate the convex hulls. The dashed lines, 0.03 eV/atom
      above the convex hulls, indicate the range in which compounds
      can be considered metastable.
    }
    \label{fig:1}
\end{figure*}

In this paper, we propose a heuristic method for such a reduced selection of prototypes based on
structure and composition analysis of related alloy families. In a previous paper~\cite{Hever2017}, we surveyed the
ICSD, focusing on O, S, and Se compounds. The database included 495 binary sulfide and 332 binary
selenide compounds. Those compounds are represented respectively by 270 and 168 different
structure types. We found considerable overlap between the sulfide and selenide structure types; i.e.,
33\% of the binary sulfide structure types also appear among binary selenides. In contrast, the overlap
with the binary oxides, which are located in the same column of the periodic table, is only about 10\%
in both cases. Here, we apply the insights from this structural analysis to look for new stable and
metastable materials in these families of materials. First, we demonstrate the prediction of new
compounds by enriching the sulfur(selenium) compound family, $A_x$S$_y$($A_x$Se$_y$), with binary structure types of selenium(sulfur), $A_x$Se$_y$($A_x$S$_y$), and oxygen, $A_x$O$_y$. Then, we further check searches based on enriching the $A_x$S$_y$($A_x$Se$_y$) families by substituting the $A$-elements in structure types of compounds that
contain sulfur or selenium with various neighboring or distant $A$-atoms in the periodic table.

\begin{figure*}[htb!]
    \centering
    \includegraphics[width=1\textwidth]{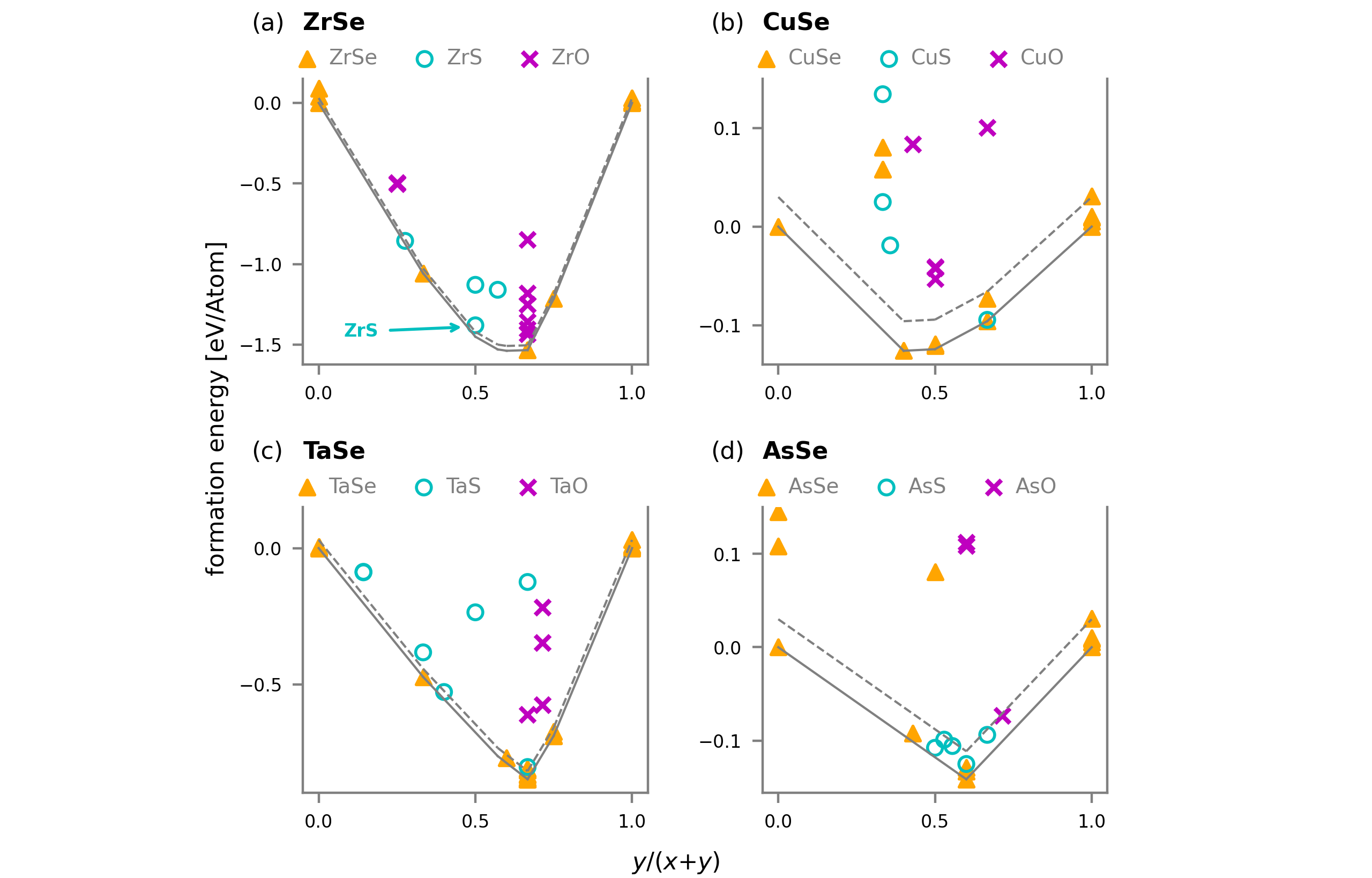}
    \caption{\small Convex hulls of the binary selenide ($A$Se) systems enriched with structure types of the corresponding binary sulfide ($A$S) and oxide ($A$O) compounds. The markers represent different prototypes, as in Figure \ref{fig:1}. 
    The black line marking the convex hull in (a) is slightly lower
    than indicated by the corresponding compounds near $x/(x+y)=0.5$
    because it includes additional new compounds found at a later
    search stage. The ZrSe compound (ZrS prototype) found in this
    stage is superseded at a later stage by a slightly more stable
    structure of the same composition (see Figure \ref{fig:3}).  }
  \label{fig:2}
\end{figure*}

This study shows that the structure types of compounds that contain certain neighboring atoms
should be much better candidates for new stable or metastable compounds. Such searches may {\it a priori} discard structure types that include distant atoms. This approach may dramatically reduce the
number of calculations required for a comprehensive search of the relevant structure space. It may
be used as a stand-alone search methodology or as an efficient starting point for the more elaborate
methods mentioned above.
\begin{figure*}[htb!]
    \centering
    \includegraphics[width=1\textwidth]{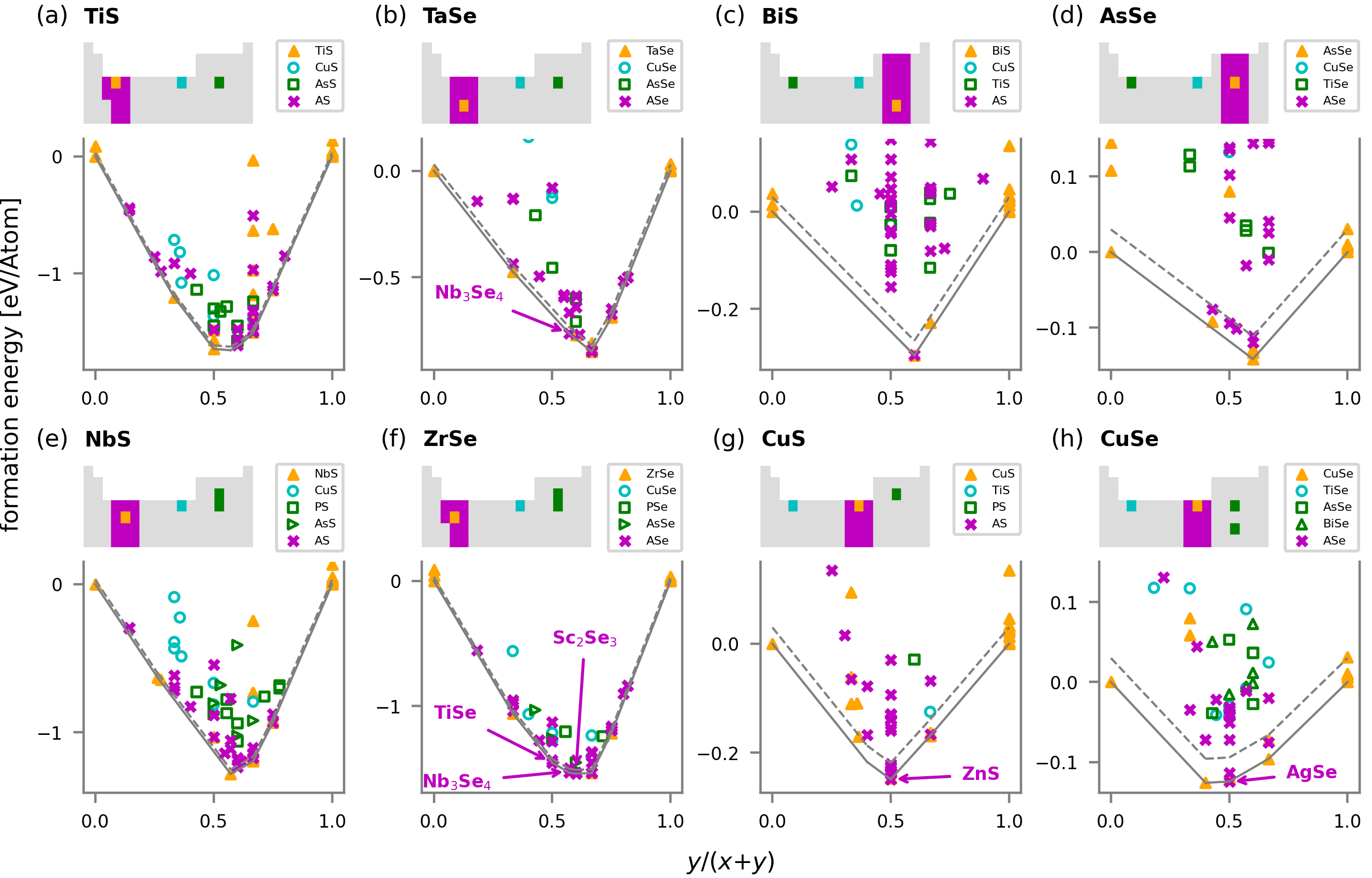}
    \caption{\small Near neighbors' enrichment, i.e., $A$S ($A$Se)
      structure types with $A$-elements from the same or adjacent
      columns. The orange triangles are the known ICSD compounds, the
      purple $x$-markers indicate the near-neighbor structures, and the
      green and cyan colors indicate $A$-elements from clusters in
      the periodic table that were found to be stoichiometry-rich in
      the sulfide and selenide binary systems~\cite{Hever2017}. The
      locations of the different $A$-atoms in the periodic table are
      marked by the corresponding colors in the inset in each frame.}
    \label{fig:3}
  \end{figure*}

\section*{Methods}

\noindent{\bf Definition of the search domain}\\
In this paper, we use a structure suggestion method based on an analysis, introduced in a previous
work~\cite{Hever2017}, of known structures included in the ICSD. The method narrows the wide variety of structure
types in the ICSD into smaller but still considerable groups of structures. As a preparatory step, we
exclude all entries with partial or random occupation and those without robust structure data. The
remaining set of entries was filtered using the AFLOW software~\cite{Curtarolo2012a}, which uses an error-checking
protocol to verify that the reported structure type, which is defined by its stoichiometry, space-group, and Pearson symbol, is consistent with the actual structure. From
the remaining list of prototypes, we focused on those that appear in binary compounds with one
element in common. The second element was selected from one of three domains of the periodic
table, inspired by our previous analysis of the ICSD~\cite{Hever2017}. This analysis showed that binary sulfides and
binary selenides exhibit a considerable proportion of shared structure types and identified
stoichiometries and structures that are present in one group but missing from the other. A smaller but
still significant overlap was found between the sulfides and selenides, on the one hand, and oxides,
on the other. This suggests the potential existence of structures that have not yet been identified in the
sulfide and selenide families. Following this reasoning, we first examined eight systems of binary
sulfides ($A_x$S$_y$, where $A$ = Cu, Nb, Ti, and Bi) and selenides ($A_x$Se$_y$, where $A$ = Cu, As, Ta, and Zr), in
which the overlap in stoichiometries, found in Ref.~\onlinecite{Hever2017}, was the smallest. The search involved enriching
each of the sulfide binaries, $A_x$S$_y$, with missing prototypes from the corresponding selenide and oxide
binaries, $A_x$Se$_y$ and $A_x$O$_y$, and each of the selenide systems by structures from the corresponding
sulfides and oxides.

In Ref.~\onlinecite{Hever2017}, it was also found that the distribution of sulfide and selenide binary structures is not uniform
across the periodic table, but tends to concentrate in three different clusters of alloying elements.
Therefore, in the second step, we extended the search to prototypes from binary systems with the elements
that are either a neighbor of the original $A$-element or belonging to these clusters.

In the third step, we further extended the search to include $A$-elements from the entire periodic table
for three of these systems, Cu$_x$S$_y$, Ti$_x$S$_y$, and Ta$_x$Se$_y$. Finally, we also examined the Ti$_x$S$_y$ system, with
prototypes suggested from titanium binary systems where the sulfur is exchanged for other elements.
This gradual extension of the search domain is aimed at comparing the efficiency of using only prototypes from
neighboring systems versus searching prototypes spanning the entire periodic table.
\\
\\

\noindent{\bf Energy and convex hull calculations}\\
The energy calculations were performed using the Vienna Ab-Initio Simulations Package (VASP)~\cite{Kresse1996, Kresse1996a}
implementation of density functional theory (DFT) with the semi-local meta-GGA functional SCAN~\cite{Sun2015}.
Each structure was fully relaxed for optimized cell shape, volume, and atom locations. The relaxation
stopped when the change in the cell's total energy was below $10^{-5}eV$. The energy of each ionic
relaxation step was calculated self-consistently with an energy convergence criterion of $10^{-6}eV$. We
employed the projector augmented wave (PAW)~\cite{Blochl1994} pseudopotentials with an energy cutoff 1.4 larger
than the VASP default (as suggested in the AFLOW standard~\cite{Calderon2015}). The k-points were sampled evenly
according to the Monkhorst-Pack method~\cite{MonkhorstPack}, and their density increased from 6000 k-points per
reciprocal atom (NKPRA) until convergence of the energy within 0.005 eV/atom.

The convex hull of each of the binary systems is constructed from the formation enthalpies of the
calculated structures, which include the ICSD's structures reported for each system and those from
the search domains. At zero temperature and pressure, the enthalpy is equivalent to the Gibbs free
energy and the total energy obtained from the DFT calculations, thus determining the thermodynamic
stability of the structures. The nodes of the convex-hull mark the stable compounds in each system.
Metastable structures lie slightly above it, within the temperature range of metallurgical processes.
The convex hull constructions were carried out by SciPy~\cite{2020SciPy-NMeth} scripts and plotted using matplotlib~\cite{Hunter2007}
Python~\cite{Oliphant2007} libraries.

\section*{Results}

\noindent{\bf Minimal Search Domain – corresponding sulfide, selenide, and oxide prototypes}\\
In our search for new binary sulfide and selenide compounds, we examined different structure types
from other compound domains. We followed several steps, gradually expanding the search domains
as described in the methods section. In the first search, we enriched the investigated sulfide (selenide)
convex hull calculations by structure types from the corresponding selenide (sulfide) and oxide systems
(i.e., the same $A$-atom but with S replaced by Se, and vice versa, or by O atoms). The convex hulls
obtained in this search are shown in Figures 1 and 2.

Figure 1 shows the calculated convex hulls of four sulfide binary systems $A$S ($A$ = Ti, Cu, Nb, and Bi).

The $A$-elements were selected following the structural analysis of ref.~\cite {Hever2017}, which indicates a relatively 
 
lower overlap of known structures in these systems compared to other pairs of selenide and sulfide
systems, and therefore presumably a higher chance of structural enrichment from the selenides. We
found two new compounds not previously listed in the ICSD: Ti$_3$S$_4$ and Cu$_3$S$_2$. In addition, ten new
compounds were found in the metastable range of 0.03 eV/atom ($\sim$~300K) above the convex hull. The new compounds belong to selenide structure types. Structure types from the binary oxide systems
led to no new stable compounds and only one of the meta-stable compounds. This result is consistent
with the analysis of Ref.~\onlinecite{Hever2017} that showed a relatively minor 10\% overlap between oxide structure types
and sulfide or selenide structure types, compared to the substantial 33\% overlap of the sulfides and
selenides.

In Figure 2, we show the convex hulls of the four selenide binary systems, $A_x$Se$_y$ ($A =$ Cu, Zr, Ta, and
As). In these systems, the minimal search approach yields one new stable compound ZrSe and nine
new meta-stable compounds not listed in the ICSD.

Overall, we tested 146 structures for these 8 systems: 98 were oxide structures that yielded no new
compounds. The 48 sulfides and selenide structures lead to the prediction of 3 new compounds, a high
discovery rate of 16 structure calculations per new compound.
\\
\\

\noindent{\bf First Extension of the Search Domain – neighboring binary systems}\\
The second domain that was examined included binary systems of ``near neighbor" elements. For
every system, $A$S ($A$Se), we define the near neighbors as $A$-elements from the same or the neighboring
columns of the periodic table (for example, for TiS, the neighbors would be $A$-elements from columns
3, 4 and 5). This extension includes 204 near-neighbor structure types added to the search described
in the first domain. In addition to these near neighbors, we included in this step several more remote
elements from areas in the periodic table that were found to be particularly stoichiometry-rich in the
structural analysis of Ref. \onlinecite{Hever2017}.

The search in this extended domain yielded five new stable compounds from the neighboring systems,
as shown in Figure 3: Ta$_3$Se$_4$ in the Nb$_3$Se$_4$ structure, CuSe in the AgSe structure, and three additional new compounds, ZrSe, Zr$_3$Se$_4$, and Zr$_2$Se$_3$, in the TiSe, Nb$_3$Se$_4$, and Sc$_2$Se$_3$ structures, respectively. The new ZrSe structure found in this step is slightly more stable and supersedes the one found in the previous step in the same stoichiometry. The discovery rate for this step, using near-neighbor structures is thus nearly one new compound per 41 calculations. Considerably lower than in the first
search, but still significant. In addition, one compound, CuS in the ZnS structure (ICSD\# 107131,
Pearson symbol hR16, space group 160), was found to be nearly degenerate (energy difference of
2 meV/atom) with the known CuS structure (ICSD\# 67581, Pearson symbol hP12, space group 194).
37 new metastable compounds were also found with near neighbors structures. The prototypes
from the stoichiometry-rich clusters, which are not near neighbors, did not yield any new compounds
or metastable structures.

\begin{figure*}[htb!]
    \centering
    \includegraphics[width=1\textwidth]{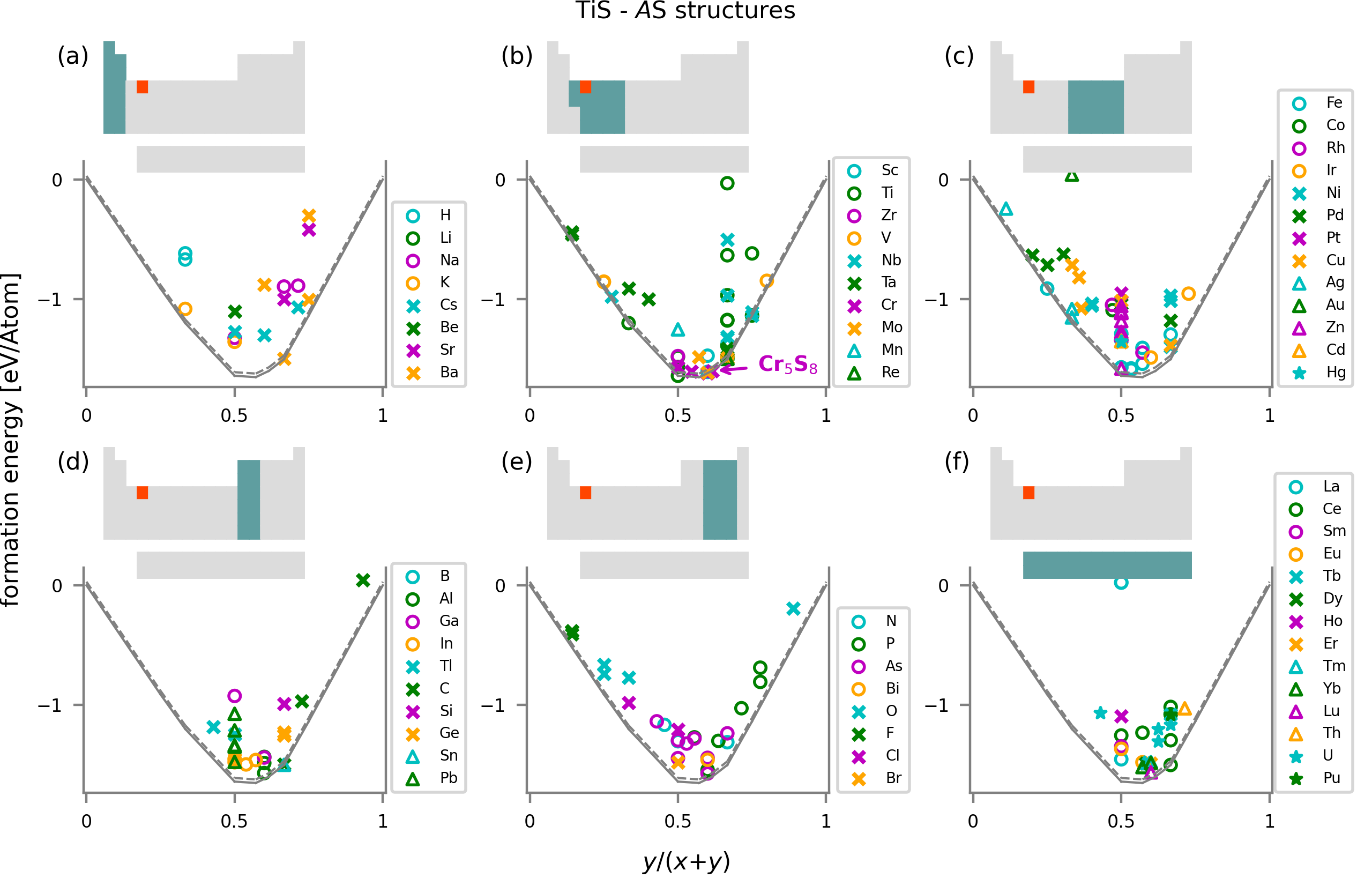}
    \caption{\small The convex hull of the TiS system enriched with structure
      types from all binary sulfide compounds $A_x$S$_y$. The six
      panels correspond to $A$-atoms from different regions of the
      periodic table. The red square in each legend marks the location
      of Ti in the periodic table and the blue region covers the
      corresponding replacement elements.}
    \label{fig:4}
  \end{figure*}

\begin{figure*}[htb!]
    \centering
    \includegraphics[width=1\textwidth]{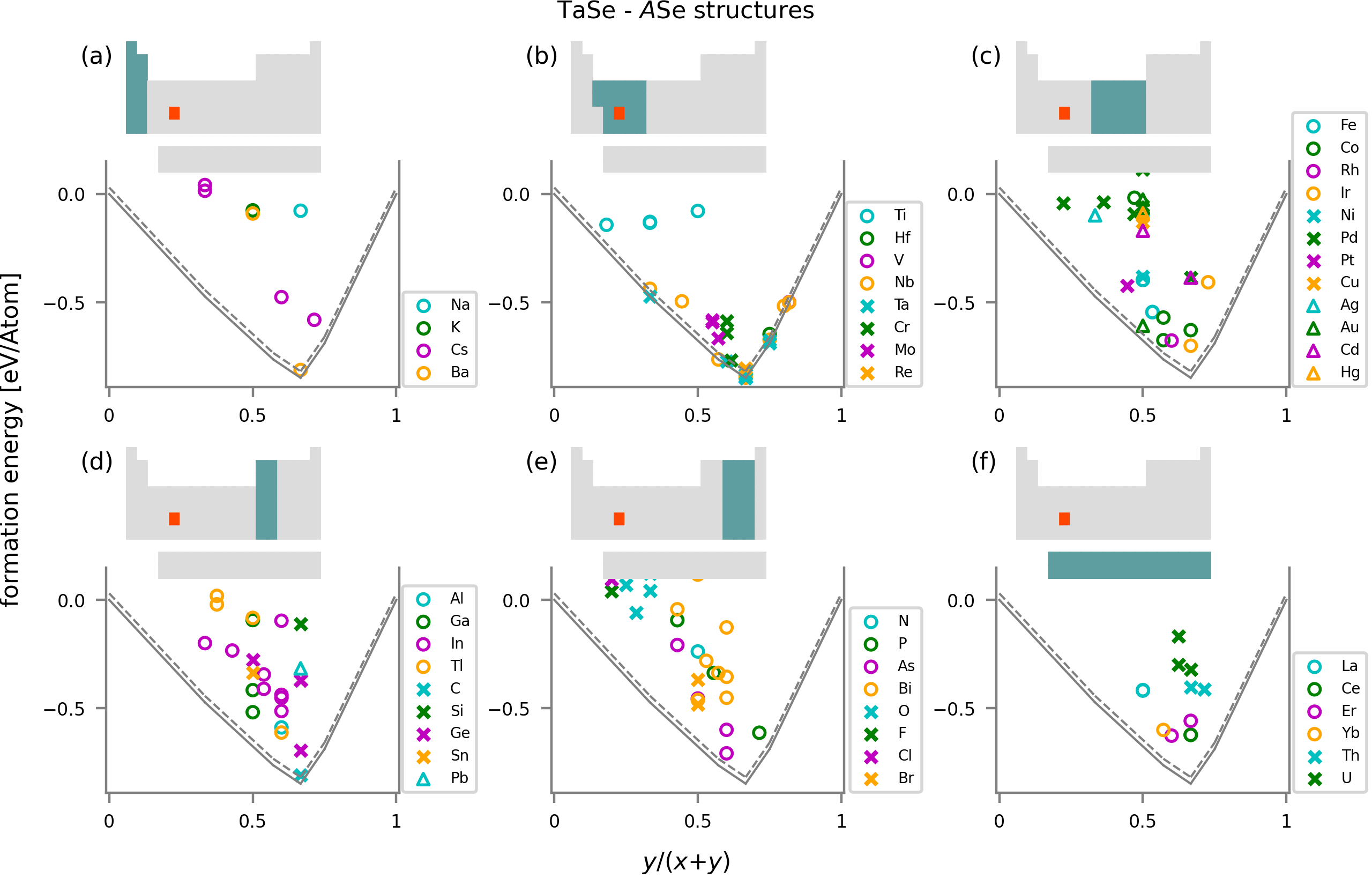}
    \caption{\small The convex hull of the TaSe system. The notations are as
      in Figure \ref{fig:4}.}
    \label{fig:5}
  \end{figure*}

\begin{figure*}[htb!]
    \centering
    \includegraphics[width=1\textwidth]{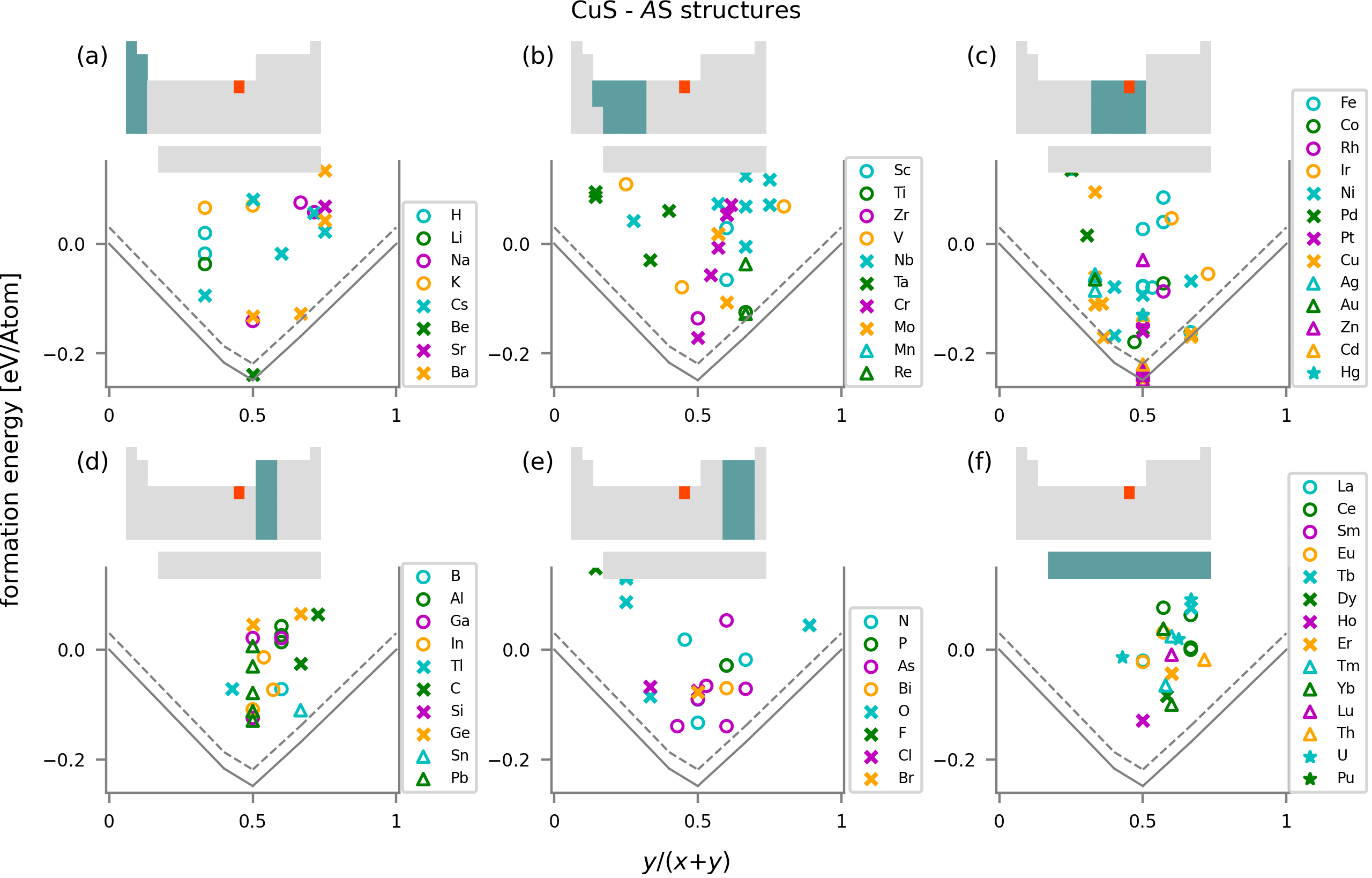}
    \caption{\small The convex hull of the CuS system. The notations are as
      in Figure \ref{fig:4}.}
    \label{fig:6}
  \end{figure*}

  \noindent{\bf Second Extension of the Search Domain}\\
  In the third step of the search, we extended the search domain to the whole periodic table; i.e., for a
  given $A$S (or $A$Se) system, we included prototypes from binary sulfide (or selenide) systems with all $A$-elements. Due to the much larger set of calculations, we limited this step to just three systems instead of the eight included in the previous ones. This step included 174, 150, and 112 prototypes for the TiS, CuS, and TaSe systems, respectively. The results are shown in Figures \ref{fig:4}-\ref{fig:6}. Each figure is a sixfold panel, where different parts show selections of $A$-elements from different regions of the periodic table. Structures common to several systems, with different $A$-atoms, are calculated once but attributed to all the corresponding systems. Expanding the search domain from the near neighbors to the rest of the table yielded only one additional compound to those identified in the previous steps, Ti$_5$S$_8$, Figure \ref{fig:4}(b). It was found in the Cr$_5$S$_8$ structure, where Cr is located in a next nearest column from Ti. This extended domain also contributed 13 metastable compounds.

\begin{figure*}[htb!]
  \centering
  \includegraphics[width=1\textwidth]{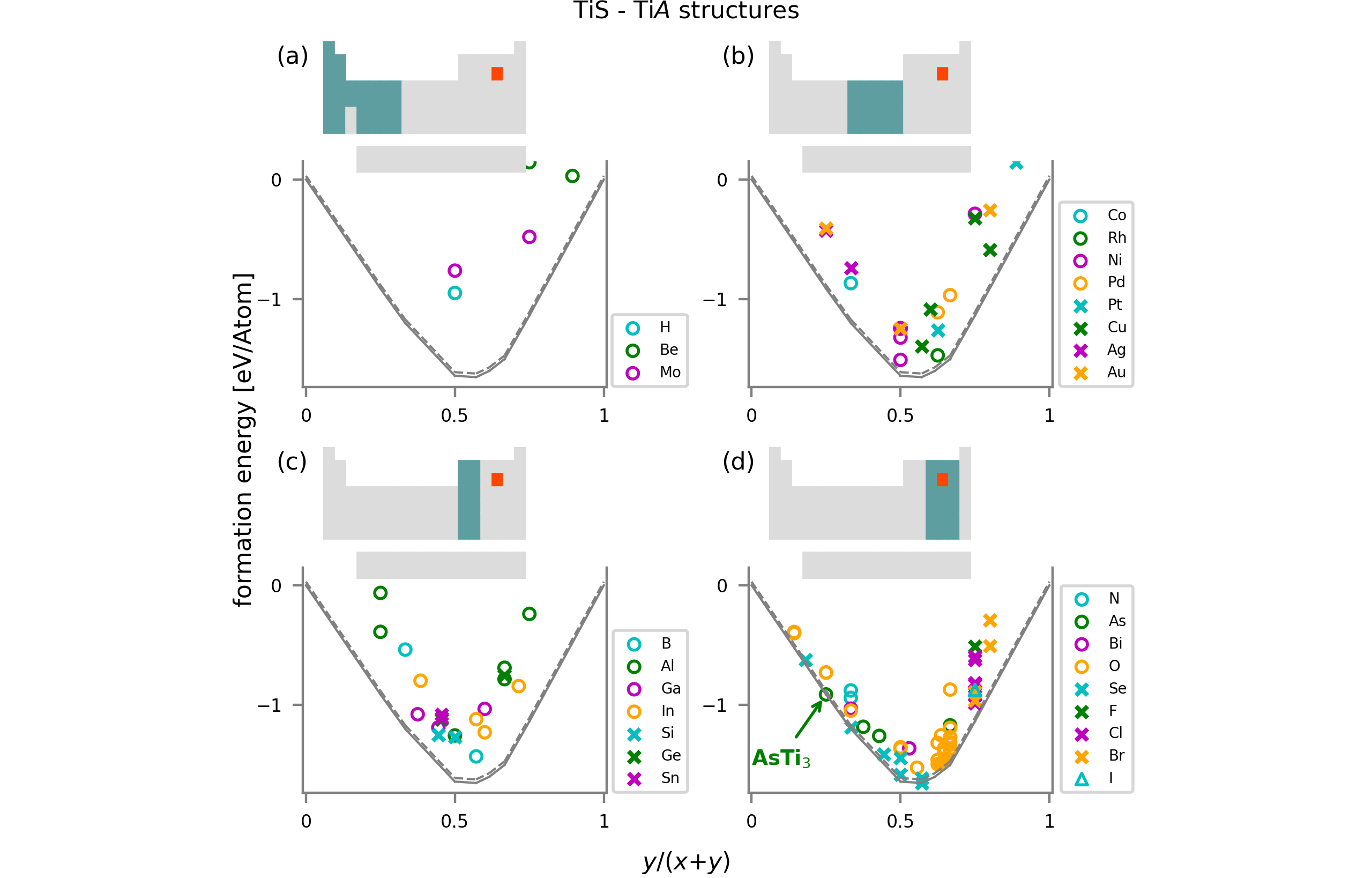}
  \caption{\small The convex hull of the TiS system enriched with
    structure types from all binary titanium compounds Ti$A$. The
    results are divided into six panels, each covering replacement
    elements from a different area of the periodic table. The red
    square marks the element S. The Lanthanides and Actinides did
    not yield any stable or metastable structures and are not shown.}
  \label{fig:7}
\end{figure*}

\noindent{\bf An alternative extension – replacing the chalcogenide element}\\
An alternative extension scheme, not based on our previous structure-types analysis of Ref. \onlinecite{Hever2017}, may
involve replacing the chalcogenide element. We applied this extension to the TiS system by including
73 structure types from binary titanium systems Ti, rather than binary sulfides. Figure \ref{fig:7} shows that
this scheme yields one additional new compound, Ti$_3$S, in the structure of Ti$_3$As, where As belongs to
a neighboring column to the chalcogenides. Non-neighbor replacements yield no new compounds,
and a trend is quite evident that further away replacements yield less stable structures.

\section*{Discussion}

Following the exhaustive calculations presented in the previous sections, we can now examine how
the different definitions of the search domain influence the search for new compounds. Overall, we
found in this study nine new compounds not previously reported in the ICSD (Table \ref{tab:1}) and one new
structure degenerate with a known compound (denoted by parentheses in Table 1). Two of these new
compounds were found in the first, most restrictive, search, out of just 48 trial structures. Five new
compounds were found in the second search domain, out of 205 trial structures borrowed from
systems with nearest columns $A$-atom replacements. The third, more extensive, search domain
yielded just one new compound out of 174 trial structures, with a structure borrowed from a system
where the $A$-atom is next nearest to the original one (Cr exchanged for Ti). The consecutive search
domains become less fruitful in the search for new compounds as they are expanded further beyond
the closest replacement elements.

\begin{table*}[htb!]
\begin{scriptsize}

\centering   

\begin{tabular}{{|l|l|l|l|l|l|l|l|l|}}
\hline
\makecell[l]{\textbf{No.} \\\textbf{}} &
\makecell[l]{\textbf{Compound}\\\textbf{}} &
\makecell[l]{\textbf{y/(x+y)}\\\textbf{}} &
\makecell[l]{\textbf{Ef}\\\textbf{[eV/atom]}} &
\makecell[l]{\textbf{Search}\\\textbf{domain}} &
\makecell[l]{\textbf{Source}\\\textbf{compound}} &
\makecell[l]{\textbf{ICSD}\\\textbf{Id}} &
\makecell[l]{\textbf{Pearson}\\\textbf{}} &
\makecell[l]{\textbf{Space}\\\textbf{group}} \\
\hline
1  & \textbf{Cu$_3$S$_2$}   & \textbf{0.4}   & -0.216  & \#1-CuSe   & Cu$_3$Se$_2$   & 239     & tP10 & 113 \\ \hline
2  & \textbf{Ti$_3$S$_4$}   & \textbf{0.571} & -1.654  & \#1-TiSe   & Se$_4$Ti$_3$   & 79629   & hP14 & 176 \\ \hline
3  & (CuS)                  & (0.5)          & (-0.249)& \#2-$A$S   & SZn            & 107131  & hR16 & 160 \\ \hline
4  & CuSe                   & 0.5            & -0.124  & \#2-$A$Se  & AgSe           & 52601   & cF8  & 216 \\ \hline
5  & \textbf{Ta$_3$Se$_4$}  & \textbf{0.571} & -0.763  & \#2-$A$Se  & Nb$_3$Se$_4$   & 16278   & hP14 & 176 \\ \hline
6  & \textbf{ZrSe}          & \textbf{0.5}   & -1.451  & \#2-$A$Se  & SeTi           & 43615   & oP8  & 62  \\ \hline
7  & \textbf{Zr$_3$Se$_4$}  & \textbf{0.571} & -1.530  & \#2-$A$Se  & Nb$_3$Se$_4$   & 16278   & hP14 & 176 \\ \hline
8  & \textbf{Zr$_2$Se$_3$}  & \textbf{0.6}   & -1.538  & \#2-$A$Se  & Sc$_2$Se$_3$   & 651804  & oF80 & 70  \\ \hline
9  & \textbf{Ti$_5$S$_8$}   & \textbf{0.615} & -1.602  & \#3-$A$*S  & Cr$_5$S$_8$    & 626594  & mS52 & 12  \\ \hline
10 & \textbf{Ti$_3$S}       & \textbf{0.25}  & -0.913  & Ti$A$*      & AsTi$_3$       & 611498  & tP32 & 86  \\ \hline
\end{tabular}

\caption{\small
New predicted stable compounds not included in the ICSD. The chemical formulas and
stoichiometries are shown (2\textsuperscript{nd} and 3\textsuperscript{rd} columns). 
New stoichiometries in the corresponding system are denoted in bold. 
$E_f$ is the formation energy, the ``search domain'' denotes the search extension step,
where $A$ denotes nearest neighbor elements and $A$* all elements of the periodic table.
The ``Source compound'' denotes the original prototype from the ICSD. 
The structure in parentheses is a new structure which is energy degenerate with a known structure.
}
\label{tab:1}

\end{scriptsize}
\end{table*}

An alternative look at the association of new compounds with near element replacement structure
types is demonstrated in the case of the TiS system, where we carried out the most comprehensive
domain search extension. The first domain included 7 and 27 structures from the Ti$_x$Se$_y$ and Ti$_x$O$_y$ 
systems, respectively, and yielded one new compound. The second domain, near neighbor $A$-replacement, included 22 $A$S structures but yielded no new compounds. The search over all $A$S
systems, with all $A$-elements, included 174 structures and yielded one new compound, with the
structure of Cr$_5$S$_8$, where Cr is in the second nearest column to Ti. Finally, the examination of 73 additional
structure types from TiA binary systems, with all $A$-elements except S, Se, and O, 17 of which are near
neighbor structure types, lead to one new compound Ti3S, in the Ti$_3$As structure, where As is a
neighbor of S. Thus, the three new compounds found in this extensive search belong to prototypes already
known in binary systems where either element is the nearest or next nearest neighbor of the original
element. The search in the first domain of the corresponding selenide structure types was the most
efficient in terms of the ratio of structures examined to compounds found.

\begin{figure*}[htb!]
    \centering
    \includegraphics[width=1\textwidth]{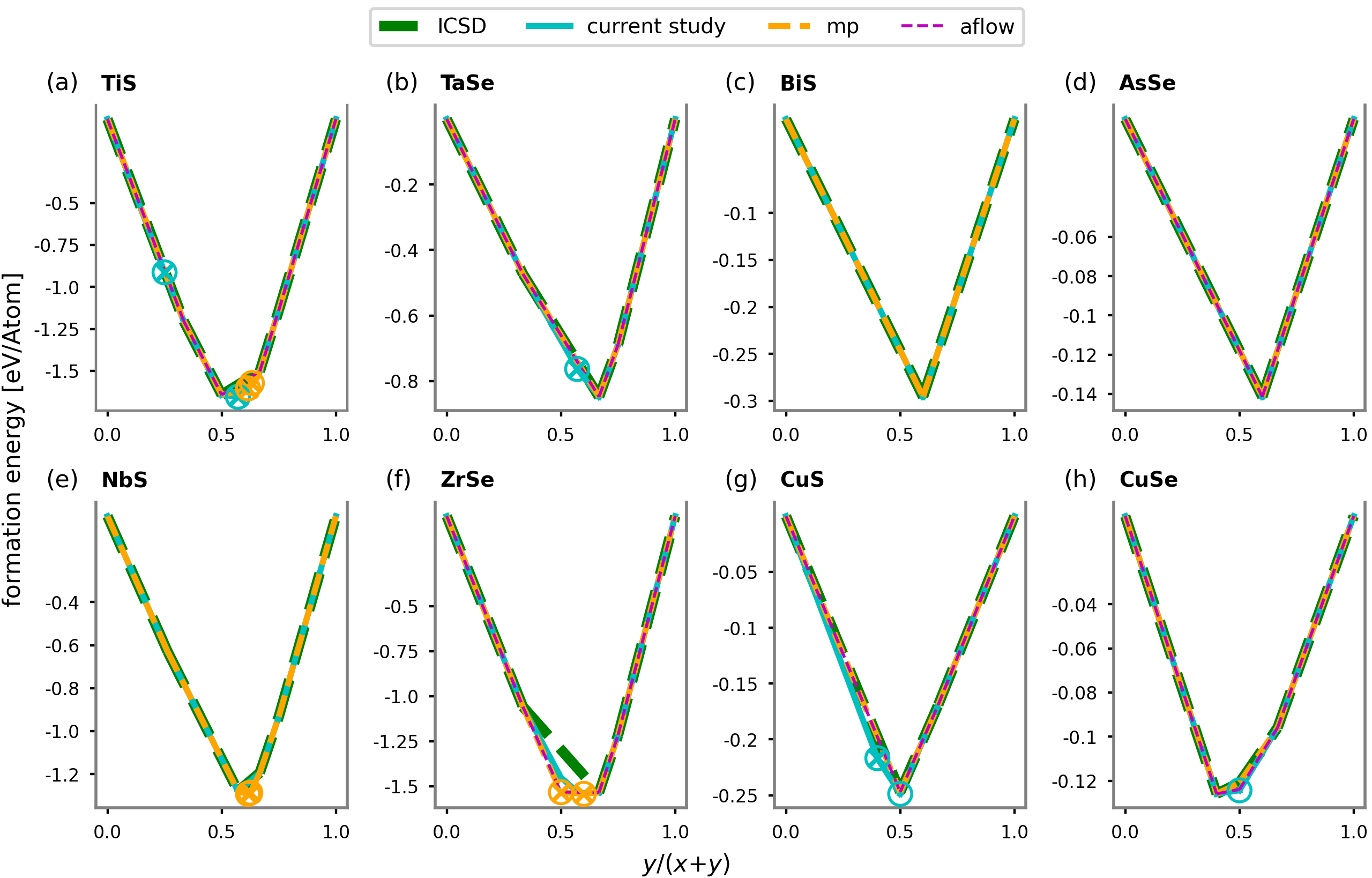}
    \caption{\small The convex hulls of the eight binary systems
      investigated in this study, including compounds from the ICSD
      only (green), MP (dashed orange), AFLOW (dashed magenta), and
      new compounds from this study (cyan). The AFLOW database does
      not include convex hulls for the BiS and NbS systems. The cyan
      and orange circles indicate compounds found in this study or
      included in the MP database, respectively. The x-marker inside a
      circle denotes an addition of a new stoichiometry to the convex
      hull, compared to the ICSD data.}
    \label{fig:8}
\end{figure*}

Figure \ref{fig:8} presents a comparison of the results of the current study with data included in two of the
most extensively used computational materials databases, Materials Project (MP)~\cite{Jain2013} and AFLOW~\cite{Curtarolo2012}. It
shows the convex hulls constructed using only structures reported in the ICSD, with convex hulls that
include the new compounds predicted in the current study, and those constructed using data from
the two databases for the same systems. The energies of structures in MP and AFLOW were obtained
using the PBE (or PBE+U) functionals. For this comparison, they were recalculated with the SCAN functional.
Table \ref{tab:2} summarizes the results from the inclusion of those structures. Adding
structures from MP and AFLOW yielded one new compound, ZrSe, which originated from a neighbor
structure type that appears in the ZrTe binary system. It is slightly lower in energy than the structures
found in the current study (of ZrS and TiSe structure types) in the same stoichiometry. Five additional
structures that appear in the MP are identified as new compounds but without an identifiable ICSD
source structure. They might have been changed due to relaxation or originated from other sources.
Some of these new structures supersede the compounds found in our study. In particular, the Ti$_5$S$_8$ compound 
(Cr$_5$S$_8$ structure) found in the third search domain is now 2meV above the MP structure that is on the
convex hull. Thus, after including all these structures, we find that, compared to the ICSD, six new
stable compounds were included in the MP database and six were uncovered in the current study.

Quantitatively, it isn't easy to precisely compare the search efficiency of our study with those
conducted to identify the compounds included in the MP and AFLOW databases. These databases
store the results of extensive calculations that usually include a few hundred structures per binary
system. Nevertheless, qualitatively, it is quite clear that a directed search, as suggested in this study,
which on average includes an order of magnitude smaller number of structures per system, should be
potentially more efficient and considerably reduce the computational cost of finding new compounds
or identifying new stoichiometries where such compounds may be found. In addition, it is also
apparent that efficient directed searches should be limited to similar systems, e.g. the replacement of
S with Se and vice versa, as shown in the first search domain of this study, or to nearest columns
replacement elements, as in the second search domain. The yield of stable structures from further
replacements is negligible.

\begin{table*}[htb!]
\begin{scriptsize}
\centering
\begin{tabular}{|l|l|l|l|l|l|l|l|}  
\hline
\makecell[l]{\textbf{No.} \\\textbf{}} &
\makecell[l]{\textbf{Compound}\\\textbf{}} &
\makecell[l]{\textbf{y/(x+y)}\\\textbf{}} &
\makecell[l]{\textbf{Ef}\\\textbf{[eV/atom]}} &
\makecell[l]{\textbf{Search}\\\textbf{domain}} &
\makecell[l]{\textbf{Id}\\\textbf{}} &
\makecell[l]{\textbf{Pearson}\\\textbf{}} &
\makecell[l]{\textbf{Space}\\\textbf{group}} \\
\hline

1 & \textbf{Cu$_3$S$_2$} & \textbf{0.4} & -0.216 & \#1\_CuSe & ICSD \# 94687 & tP10 & 113 \\ \hline
2 & \textbf{Ti$_3$S$_4$} & \textbf{0.571} & -1.654 & \#1\_TiSe & ICSD \# 79629 & hP14 & 176 \\ \hline
3 & CuS & 0.5 & -0.249 & \#2\_AS & ICSD \# 107131 & hR16 & 160 \\ \hline
4 & CuSe & 0.5 & -0.124 & \makecell{\#2\_ASe \\ AFLOW} & \makecell[l]{ICSD \# 52601 \\ aflow-25e868231754fe59} & cF8 & 216 \\ \hline
5 & \textbf{Ta$_3$Se$_4$} & \textbf{0.571} & -0.763 & \#2\_ASe & ICSD \# 16278 & hP14 & 176 \\ \hline
6 & \textbf{Ti$_3$S} & \textbf{0.25} & -0.913 & TiA* & ICSD \# 611498 & tP32 & 86 \\ \hline
7 & \textbf{Ti$_5$S$_8$} & \textbf{0.615} & -1.604 & mp & mp-1208223 & mS13 & 12 \\ \hline
8 & \textbf{Ti$_7$S$_12$} & \textbf{0.632} & -1.573 & mp & mp-673657 & aP38 & 2 \\ \hline
9 & \textbf{Nb$_5$S$_8$} & \textbf{0.615} & -1.292 & mp & mp-1220654 & hP39 & 149 \\ \hline
10 & \textbf{Nb$_3$S$_5$} & \textbf{0.625} & -1.287 & mp & mp-32983 & aP32 & 1 \\ \hline
11 & \textbf{ZrSe} & \textbf{0.5} & -1.533 & mp & mp-1183040 & hP2 & 187 \\ \hline
12 & \textbf{Zr$_2$Se$_3$} & \textbf{0.6} & -1.542 & mp & mp-1215598 & mS10 & 12 \\ \hline
\end{tabular}
\caption{\small Compounds predicted after including structures from the AFLOW and MP databases. The column notations are as in Table \ref{tab:1}.}
\label{tab:2}
\end{scriptsize}
\end{table*}

\section*{Summary}

We presented a comprehensive search for new stable compounds in eight binary sulfide and selenide
systems by searching through candidate structure types screened from the ICSD. All candidate
structure types represent compounds that share one element with the explored binary system and
the second from three different domains, increasing in size, of the periodic table. The first search
domain included in each of the $A$S ($A$Se) studied systems structure types of compounds from the
corresponding $A$Se ($A$S) and $A$O systems. The second domain included compound structure types
from $A$S ($A$Se) with $A$-atoms from the same or neighbor columns. The third domain included the rest
of the periodic table elements for the $A$-atom. For the TiS system, we applied a further expansion and
searched TiA structures, with $A$-elements from all over the periodic table.

We found ten new stable compounds not reported in the ICSD in six out of the eight investigated
binary systems. Two of these compounds were found in the first search domain, six in the second, and
two in the third. Nine of the ten predicted compounds were obtained from prototypes reported in
neighboring binary systems. The compound found in the third domain was obtained from a second
neighbor prototype.

The results were further enhanced by adding structures indicated as stable in those systems from the
computational structure databases of AFLOW and the materials project (MP). As a result, the total
number of new structures relative to the ICSD was increased to twelve. Six of them are from MP and six are from the current work (one shared with AFLOW).

Our results demonstrate that limiting the search domain to structure prototypes indicated by near-neighbor systems can offer a cost-effective strategy to search for new stable compounds. This
approach can be used as a stand-alone method, as done in this study, or as a starting seed for more
elaborate search algorithms.

\section*{Acknowledgments}
AN and OL acknowledge funding from the PAZY foundation grant 233/20.

\end{document}